\documentstyle[12pt,psfig]{article}
\input xy
\xyoption{all}

\begin{document}

\hfill ILL-(TH)-03-03

\hfill hep-th/0302152

\vspace{1.0in}

\begin{center}


{\large\bf Discrete Torsion and Shift Orbifolds}

\vspace{0.75in}

Eric Sharpe \\
Department of Mathematics \\
1409 W. Green St., MC-382 \\
University of Illinois \\
Urbana, IL  61801 \\
{\tt ersharpe@uiuc.edu} \\

$\,$

\end{center}

In this paper we make two observations related to discrete torsion.
First, we observe that an old obscure degree of freedom
(momentum/translation shifts)
in (symmetric) string orbifolds is related to discrete torsion.
We point out how our previous derivation of discrete torsion from
orbifold group actions on $B$ fields includes these
momentum lattice shift phases, 
and discuss
how they are realized in terms of orbifold group actions on D-branes.  
Second, we describe the M theory dual of IIA discrete torsion.
We show that IIA discrete torsion is encoded in analogues of 
the shift orbifolds above for the M theory $C$ field.

\begin{flushleft}
February 2003
\end{flushleft}

\newpage

\tableofcontents

\newpage

\section{Introduction}

Discrete torsion is an obscure-looking degree of freedom in
string orbifolds, a degree of freedom measured by the group
cohomology group $H^2(G, U(1))$ \cite{vafa1}.
Its unusual form has, over the years, prompted numerous speculations
concerning its nature.

Recently, it was shown \cite{medt,cdt,hdt,dtrev} that discrete torsion
is simply a consequence of having a $B$ field -- discrete torsion
is an example of a degree of freedom arising from defining an
orbifold group action on any theory with a $B$ field,
and has nothing to do with CFT {\it per se}.
More generally, the orbifold group action on any field with a
gauge invariance is not uniquely determined by specifying the
group action on the base space -- one can combine any group
action on the field with a gauge transformation, to get a new
group action on the field.  Not only has it been shown that
possible group actions on a $B$ field are (partially) counted
by  $H^2(G, U(1))$, but the phase factors originally used to
define discrete torsion in \cite{vafa1} have also been derived
from the same first principles, and Douglas's conjectured
description of discrete torsion for D-branes \cite{doug1,doug2} has
also been derived from first-principles.

However, although group actions on the $B$ field include
$H^2(G, U(1))$, there are also additional group actions on $B$ fields
which are not counted by $H^2(G, U(1))$.

In the first part of this paper, we shall comment extensively
on these group actions that do not arise from ordinary discrete torsion.
Using the methods of \cite{medt,dtrev}, we shall argue that 
for toroidal orbifolds, these other degrees of freedom define
(manifestly $SL(2,{\bf Z})$-invariant) one-loop phase factors of the form
\begin{equation}   \label{base}
\exp\left( i ( p_R a_L \: - \: p_L a_R) \right)
\end{equation}
which the reader may recognize as defining momentum/winding phase factors
or ``shift orbifolds'' \cite{dgh,katrin}
whose asymmetric generalization appeared in, for example, \cite{asymm1}.
We shall see these same degrees of freedom also appear nontrivially in D-branes,
and correspond to orbifold group actions on D-brane bundles ${\cal E}$
that map ${\cal E} \mapsto {\cal E} \otimes {\cal L}$,
where ${\cal L}$ is a line bundle with connection,
whose holonomies are encoded in the $a_L$, $a_R$ above.

These momentum lattice shifts are seen most commonly in 
asymmetric orbifolds and in heterotic toroidal orbifolds,
where they correspond to orbifold Wilson lines, for example.
They also appear in symmetric orbifolds, although in symmetric
orbifolds they are much more boring, and so tend to be neglected.
In particular, in symmetric orbifolds, these shifts often 
correspond to merely changing
the radius of the torus one has compactified on \cite{dgh,katrin},
making them relatively uninteresting by themselves.
Our point is not to investigate these degrees of freedom {\it per se},
but rather to use them to clarify the physical meaning of part of our work
on discrete torsion \cite{medt,dtrev}, and, in the second part
of this paper, to understand the M theory
dual of IIA discrete torsion.

Note also that discrete torsion is sometimes incorrectly described
as the {\it only} possible modular-invariant phase factor.
Rather, discrete torsion is merely one possible modular-invariant
phase factor; other possibilities have existed implicitly in the
physics literature for many years.  (Additional possible
modular-invariant phase factors, not necessarily arising
from $B$ field group actions, also exist \cite{psapriv}.)
Momentum lattice shifts also give rise to modular-invariant phase
factors, albeit more complicated ones.
Recall that ordinary discrete torsion can be described as a 
mere multiplicative phase factor on each twisted sector contribution
to an orbifold partition function, as
\begin{displaymath}
Z \: = \: \sum_{{\scriptsize \begin{array}{c} g,h\\gh=hg \end{array}}} \epsilon(g,h) Z_{g,h}
\end{displaymath}
at one-loop.  By contrast, the phase factors $\epsilon(g,h,p_R,p_L)$
resulting from momentum lattice shift orbifolds change each
$Z_{g,h}$ internally.  For example, the one-loop partition function
of a toroidal orbifold (with shift factors) can be formally expanded
\begin{displaymath}
Z \: = \: \sum_{{\scriptsize \begin{array}{c} g,h\\gh=hg \end{array}}} \sum_{p_R,p_L} 
\epsilon(g,h,p_R,p_L)
Z_{g,h}(p_R,p_L)
\end{displaymath}
where
\begin{displaymath}
Z_{g,h} \: = \: \sum_{p_R,p_L} Z_{g,h}(p_R,p_L)
\end{displaymath}
So, the momentum lattice shift orbifolds also act in the orbifold
partition function by phases, but unlike ordinary discrete torsion,
they do not merely multiply each $Z_{g,h}$ by a constant.

Upon reflection, the existence of additional (albeit more
complicated) modular-invariant phase factors should
not surprise the reader; after all, for example, the Cartan-Leray
spectral sequence for the (equivariant) cohomology of a quotient $X/G$
gets contributions not only from $H^2(G,U(1))$, but also from $H_1(G,
H_1(X))$.  The shift orbifolds correspond, roughly,
to the second contribution. 

Also note that in principle, our observations allow us to make
sense of momentum shift phases for orbifolds of non-simply-connected
spaces other than tori, though we shall not pursue that direction here.

In the second part of this paper, 
we shall use these shift degrees of freedom to solve an old puzzle,
specifically, how to see IIA discrete torsion in M theory?
Since M theory has no $B$ field, and discrete torsion is intimately
related to $B$ fields, there appears to be a significant problem here.
We shall see explicitly that IIA discrete torsion appears in M theory as
an analogue for $C$ fields of the degree of freedom discussed above;
in particular, possible orbifold group actions on the $C$ field,
on a space compactified on $S^1$, include discrete torsion of the
compactified theory.

We begin in section~\ref{review} by reviewing the general story
of orbifold group actions on $B$ fields, as developed in \cite{medt}.
In particular, we review the additional degrees of freedom, beyond
discrete torsion, that crop up by considering orbifold group actions
on $B$ fields.
In section~\ref{phases} we discuss the twisted sector phase factors
for these additional degrees of freedom, and show that for toroidal
orbifolds, the twisted sector phase factors can be written in
the form $\exp( p_L a_R - p_R a_L)$.  In section~\ref{modinvck} we
explicitly check modular invariance of the phases, using expressions
for the phases originating from \cite{medt}.  
In section~\ref{dbranes} we discuss
the orbifold group actions on D-branes corresponding to these degrees
of freedom, by specializing the general analysis of \cite{medt}.
In section~\ref{consistency1} we perform our first independent consistency
test of our results, by repeating, for our additional degrees of freedom,
a calculation in \cite{doug1,pauldt} deriving closed-string twisted
sector phase factors from group actions on D-branes.  
In section~\ref{vwanalogue} we discuss how these degrees of freedom
act in two particular examples of closed string orbifolds, 
namely $T^4/{\bf Z}_2$ and $T^2/{\bf Z}_2$.
In section~\ref{consistency2} we perform a second consistency check
of our results, by repeating the analysis of \cite{gomis} for our
shift orbifolds.  Finally, in section~\ref{mdual} we discuss
how analogues of these additional degrees of freedom for the M theory
$C$ field provide the M theory dual to IIA discrete torsion,
as well as to the degrees of freedom for the IIA $B$ field discussed 
in this paper.

In order to try to simplify the exposition, in the first
part of this paper we shall assume the vacuum expectation
value of the $B$ field vanishes, so as to simplify
momentum lattices and hopefully make our methods more clear.
This assumption will not change the possible gauge transformations
of the $B$ field, which form the basis of our calculations.

\section{Review of group actions on $B$ fields}   \label{review}

Let us recall the result of \cite{medt}:  the difference
between two orbifold group actions on a $B$ field\footnote{
Any $B$ field will do, so long as at least one orbifold
group action exists.  Our methods do {\it not} assume the
$B$ field is flat; however, only for topologically trivial $B$ fields
will orbifold group actions always be guaranteed to exist.}
is defined by the following data:
\begin{enumerate}
\item A set of principal $U(1)$-bundles $T^g$, one for each
$g \in G$, and each possessing a flat $U(1)$ connection $\Lambda(g)$.
\item Connection-preserving maps $\omega^{g_1, g_2}: T^{g_2} \otimes g_2^* T^{g_1} \rightarrow
T^{g_1 g_2}$, such that the following diagram\footnote{
In writing diagram~(\ref{omegacocycle}), we refer to tensor products
of principal $U(1)$ bundles, which the reader has almost certainly
not seen before.  More properly, we can think about this as a tensor
product of abelian torsors, or, if the reader prefers, we can replace
the principal $U(1)$ bundles with line bundles with hermitian fiber metrics.
In either case, the result is the same:  the transition functions of
such a tensor product are the product of the transition functions of the
factors, and a connection can be formed on the tensor product via the
sum of the connections on either factor. 
} commutes:
\begin{equation}  \label{omegacocycle}
\xymatrix@C+50pt{
T^{g_3} \otimes g_3^* \left( \, T^{g_2} \otimes g_2^* T^{g_1} \right)
\ar[r]^{ \omega^{g_1, g_2}}
\ar[d]_{  \omega^{g_2, g_3} } &
T^{g_3} \otimes g_3^* T^{g_1 g_2}
\ar[d]^{ \omega^{g_1 g_2, g_3} } \\
T^{g_2 g_3} \otimes (g_2 g_3)^* T^{g_1}
\ar[r]^{ \omega^{g_1, g_2 g_3}} &
T^{g_1 g_2 g_3}
}
\end{equation}
\end{enumerate}
subject to gauge invariances of the bundles,
which descend to the maps $\omega^{g_1,g_2}$ as follows:
\begin{equation}    \label{omegacobound}
\omega'^{g_1, g_2} \: \equiv \:
\kappa_{g_1 g_2} \circ \omega^{g_1, g_2} \circ
\left( \kappa_{g_2} \otimes g_2^* \kappa_{g_1} \right)^{-1}
\end{equation}
where $\kappa_g$ defines a gauge transformation\footnote{In fact, we can replace
any bundle $T^g$ with an isomorphic bundle, with isomorphic connection,
and the $\kappa_g$'s define bundle isomorphisms in general.} of the bundle $T^g$.

We recover ordinary discrete torsion by considering the special case
that the flat bundles $T^g$ are all trivializable, with gauge-trivial
connections.  In this case, and assuming the underlying space is
connected, the connection-preserving bundle isomorphisms
$\omega^{g_1,g_2}$ are completely determined by constant maps into $U(1)$,
{\it i.e.}, elements of $U(1)$.  The constraint~(\ref{omegacocycle}) 
tells us these constants are chain cocycles, and the residual gauge
invariance~(\ref{omegacobound}) tells us to ignore coboundaries,
hence we recover degree two group cohomology.

However, not all flat bundles are topologically trivial, and not all
flat connections even on a trivial bundle are gauge-trivial.
Thus, there are slightly more degrees of freedom arising from orbifold
group actions on $B$ fields than merely discrete torsion.

These remaining degrees of freedom are the subject of this paper.
We shall see that they correspond to phase factors of the form
$\exp\left( p_L a_R - p_R a_L\right)$ in toroidal orbifolds,
and correspond to orbifold group actions on D-branes in which
the gauge bundle ${\cal E}$ is tensored with a line bundle,
and these results will be checked for consistency in several different
ways.

Of course, one can turn on both discrete torsion and these
momentum shift factors; for simplicity, we shall concentrate
on the latter phase factors only.

\section{Twisted sector phase factors}  \label{phases}

In order to understand these extra degrees of freedom arising
from orbifold group actions on the $B$ field, let us consider
the phase factors appearing in one-loop twisted sectors.

In \cite{medt} we argued that the one-loop twisted sector
phase factor for {\it any} orbifold group action on the $B$ field
was given by
\begin{equation}  \label{good1loop}
\left( \omega^{g,h}_x \right) \,
\left( \omega^{h,g}_x \right)^{-1} \,
\exp \, \left( \, \int^{hx}_x \, \Lambda(g) \: - \:
\int^{gx}_x \Lambda(h) \, \right)
\end{equation}
(omitting factors of $\pi$ and $i$ in the exponential, for convenience.)
Recall from \cite{medt} that this phase factor arose from evaluating
$\exp(B)$ in a one-loop twisted sector, {\it i.e.} with branch cuts.
The phase above arose from the $B$ field gauge transformations on the
boundaries, and is closely analogous to the notion of orbifold Wilson lines.

For ordinary discrete torsion, we took the bundles $T^g$ to be trivial,
with gauge-trivial connections $\Lambda(g)$, in which case the $\omega$'s
became constant maps.  In this case,
expression~(\ref{good1loop}) simplifies to become
merely
\begin{displaymath} 
\frac{ \omega^{g,h} }{ \omega^{h,g} }
\end{displaymath}
which the reader will recognize as being the standard expression for
discrete torsion phase factors at one-loop \cite{vafa1}.

Next, let us consider the case of toroidal orbifolds, and the phase factors
discussed in the previous section.  For these phase factors, in the
case of a torus, we can take the $\omega$ to all be the identity element
of $U(1)$, and expression~(\ref{good1loop}) becomes
\begin{equation}  \label{phstart}
\exp \, \left( \, \int^{hx}_x \, \Lambda(g) \: - \:
\int^{gx}_x \Lambda(h) \, \right)
\end{equation}
Now, let us simplify this expression.  Consider a twisted sector defined
by sigma model maps \cite[section 10.1]{lt}
\begin{eqnarray*}
X_L^i (\tau + \sigma) & = & \frac{1}{2} x^i \: + \: p_L^i (\tau + \sigma)
\: + \: \mbox{oscillators} \\
X_R^i (\tau - \sigma) & = & \frac{1}{2}x^i \: + \: p_R^i (\tau - \sigma)
\: + \: \mbox{oscillators}
\end{eqnarray*}
where $X^i(\tau,\sigma) \: = \: X_L + X_R$.
(In the first part of this paper, to simplify the exposition,
recall we assume $B \equiv 0$, so the $p_{L,R}$ above can be identified
with elements of the momentum lattice.)
The phase factor~(\ref{phstart}) can be written less formally as
\begin{displaymath}
\exp \, \left( \int_0^{2 \pi} \, \Lambda(g)_i \frac{ \partial X^i }{
\partial \sigma} d \sigma
\: - \: \int_0^{2 \pi} \, \Lambda(h)_i \frac{ \partial X^i}{
\partial \tau} d \tau \, \right)
\end{displaymath}
where $X^i(\tau=0,\sigma=0) = x$, $X^i(\tau=0,\sigma=2\pi) = hx$,
$X^i(\tau=2\pi,\sigma=0) = gx$, and $X^i(\tau=2\pi, \sigma =2\pi) = ghx$.
Since we are working with flat connections
on a toroidal target space, we can assume that
the gauge fields $\Lambda(g)$ are constant on the target:
if the symbol $d x^i$ is used to denote generators of target-space
cohomology, then we can write $\Lambda(g) = \Lambda(g)_i d x^i$ for constants
$\Lambda(g)_i$.
The phase factor~(\ref{phstart}) can now be written as
\begin{eqnarray*}
\lefteqn{ \exp \left( \Lambda(g)_i \int_x^{hx} \frac{ \partial X^i }{ \partial \sigma}
d \sigma \: - \: \Lambda(h)_i \int_x^{gx} \frac{ \partial X^i }{ \partial
\tau} d \tau \right) } \\
& = & \exp \left( \: \Lambda(g)_i \left( p_L^i - p_R^i \right) \: - \:
\Lambda(h)_i \left( p_L^i + p_R^i \right) \: \right)
\end{eqnarray*}
If we define
\begin{eqnarray*}
a_{ R i}  & = & \Lambda(g)_i  \: + \:  \Lambda(h)_i  \\
a_{L i} & = & \Lambda(g)_i  \: - \:  \Lambda(h)_i  
\end{eqnarray*}
then we finally see the phase factor~(\ref{phstart}) can be written 
in the advertised form, as
\begin{displaymath}
\exp\left( p_L^i a_{R i} \: - \: p_R^i a_{L i} \right)
\end{displaymath}
(where we have omitted various factors of $\pi$ and $i$),
which corresponds to phase factors discussed in {\it e.g.} \cite{dgh,katrin}.

Note that in principle, one could use expression~(\ref{phstart}) to make
sense of momentum shift phases in orbifolds of non-simply-connected spaces
other than tori, though we shall not do so here.

In passing, note that modular invariance of this phase factor,
in the form $\exp( i p \cdot a)$, is manifest.
Modular invariance of the expression~(\ref{phstart}) will be checked
in the next section.

We should also take a moment to compare this phase factor to the corresponding
asymmetric orbifold phase factor, as discussed in \cite{asymm1}.
Strictly speaking, the phase factor above has exactly the same form as the
one written in the introduction of \cite{asymm1},
but we are discussing symmetric orbifolds here, not asymmetric orbifolds.
However, note that the phase factor described explicitly in \cite{asymm1}
referred to the group action on an untwisted sector state;
it was the phase factor for a $(g,1)$ one-loop twisted sector,
in other words.  By contrast, the phase factor above occurs
in symmetric orbifolds in a general $(g,h)$ twisted sector.
In a $(1,g)$ twisted sector $a_L = a_R$ and so we see that
we have correctly recovered the {\it symmetric} orbifold phase factor.
The fact that the $(g,1)$ twisted sector phase factor in an asymmetric
orbifold, has the same form as the $(g,h)$ twisted sector phase factor
in a symmetric orbifold, is merely a coincidence.

\section{Explicit verification of modular invariance}  \label{modinvck}

Let us take a moment to verify modular invariance 
explicitly for these phase factors $\exp( i p \cdot a)$, at string one-loop,
written in the form of 
expression~(\ref{phstart}).

Under the following element of $SL(2, {\bf Z})$:
\begin{equation}    \label{gnlsl2z}
\left( \begin{array}{cc}
       a & b \\
       c & d
       \end{array} \right)
\end{equation}
the group elements $(g,h)$ transform as a doublet:
\begin{displaymath}
\left( \begin{array}{c}
       g \\ h \end{array} \right) \: \mapsto \:
\left( \begin{array}{c}
       g^a h^b \\ g^c h^d \end{array} \right)
\end{displaymath}
(Note that in one-loop twisted sectors the group elements
$g$ and $h$ commute, so there is no ordering ambiguity.)

Ordinary discrete torsion is invariant under this transformation.
If we define
\begin{displaymath}
\epsilon(g,h) \: = \: \frac{ \omega^{g,h} }{ \omega^{h,g} }
\end{displaymath}
to be the one-loop phase factor defining discrete torsion,
then $\epsilon(g,h) = \epsilon(g^a h^b, g^c h^d)$.

A similar statement is true for the phase factors
$\exp \left( p_L a_R - p_R a_L\right)$, as we shall outline by examining
them expressed in the form~(\ref{phstart}).

We shall first verify modular invariance in detail for the
matrix
\begin{equation}  \label{firstsl2z}
\left( \begin{array}{cc}
       1 & 1\\
       0 & 1
       \end{array} \right)
\end{equation}
The computation will require being careful about pullbacks,
and is messy in general.  After we have worked out this case
in detail, we shall outline the more general case, but for
simplicity of exposition will omit pullbacks when doing the
general verification.

Acting with the $SL(2,{\bf Z})$ element~(\ref{firstsl2z}),
the one-loop phase factor $\exp( i p \cdot a)$, written in the form
of expression~(\ref{phstart}), becomes
\begin{displaymath}
\exp \left( \, \int_x^{hx} \Lambda(hg) \: - \:
\int_x^{ghx} \Lambda(h) \, \right)
\end{displaymath}
The second integral, from $x$ to $ghx$, proceeds along the
path $x \rightarrow gx$, then along the path $gx \rightarrow ghx$.
Thus, we can write the second integral as
\begin{eqnarray*}
\int_x^{ghx} \Lambda(h)  & = & \int_x^{gx} \Lambda(h) \: + \:
\int_{gx}^{ghx} \Lambda(h) \\
 & = & \int_x^{gx} \Lambda(h) \: + \:
\int_x^{hx} g^* \Lambda(h)
\end{eqnarray*}
To evaluate the first integral, we use the identity
\begin{displaymath}
\Lambda(hg) \: = \: \Lambda(g) \: + \: g^* \Lambda(h)
\end{displaymath}
(Since the $\Lambda(g)$'s are $U(1)$ gauge fields, they are only equivalent up
to a gauge transformation, which turns out to encode ordinary
discrete torsion.  Since we are assuming there is no ordinary discrete torsion,
and concentrating solely on the momentum shift phase orbifolds,
those gauge transformations are trivial, and so we are ignoring their
contribution.)
It is now straightforward to check algebraically, using the
expressions above, that offending terms cancel and
\begin{displaymath}
\exp \left( \, \int_x^{hx} \Lambda(hg) \: - \:
\int_x^{ghx} \Lambda(h) \, \right)
\: = \:
\exp \left( \, \int_x^{hx} \Lambda(g) \: - \:
\int_x^{gx} \Lambda(h) \, \right)
\end{displaymath}
so our phase factor $\exp( i p \cdot a)$ is $SL(2,{\bf Z})$-invariant
in this case.

Next, we shall outline how to check $SL(2,{\bf Z})$ invariance for
general $SL(2,{\bf Z})$ elements.
Carefully keeping track of pullbacks in the general calculation
is straightforward but messy, so for the purposes of simplifying the 
present exposition, we shall drop pullback symbols.

Dropping pullback notations for convenience, under an $SL(2,{\bf Z})$
transformation~(\ref{gnlsl2z}) the phase factor $\exp( i p \cdot a)$,
written in the form of expression~(\ref{phstart}) becomes
\begin{displaymath}
\exp \left( \int_x^{g^c h^d x} \Lambda(g^a h^b) \: - \:
\int_x^{g^a h^b x} \Lambda(g^c h^d) \right)
\end{displaymath}
Dropping pullback notations, we can write
\begin{eqnarray*}
\Lambda(g^ah^b) & = & a \Lambda(g) \: + \: b \Lambda(h) \\
\Lambda(g^ch^d) & = & c \Lambda(g) \: + \: d \Lambda(h) 
\end{eqnarray*}
as well as
\begin{eqnarray*}
\int_x^{g^a h^b x} \Lambda(g^c h^d) & = &
a \int_x^{gx} \Lambda(g^c h^d) \: + \: b \int_x^{hx} \Lambda(g^ch^d) \\
\int_x^{g^ch^dx} \Lambda(g^a h^b) & = & 
c \int_x^{gx} \Lambda(g^a h^b) \: + \:
d \int_x^{hx} \Lambda(g^a h^b)
\end{eqnarray*}
It is now a simple matter of algebra to verify that
\begin{eqnarray*}
\exp \left( \int_x^{g^c h^d x} \Lambda(g^a h^b) \: - \:
\int_x^{g^a h^b} \Lambda(g^c h^d) \right)
& = &
\exp \left( (ad-bc) \int_x^{hx} \Lambda(g) \: - \:
(ad-bc) \int_x^{gx} \Lambda(h) \right) \\
& = & \exp \left( \int_x^{hx} \Lambda(g) \: - \:
\int_x^{gx} \Lambda(h) \right)
\end{eqnarray*}
Thus, we see that the one-loop $(g,h)$ twisted sector
phase factor $\exp( i p \cdot a)$,
written in the form of expression~(\ref{phstart}),
is $SL(2,{\bf Z})$-invariant.

\section{Action on D-branes}  \label{dbranes}

Next, let us consider induced actions on D-branes in orbifolds.

In general terms, orbifold group actions on the $B$ field necessarily
mix with orbifold group actions on D-branes because of the relationship
between gauge transformations of the $B$ field and affine translations
of the Chan-Paton factors.  As is well-known, for consistency,
under a gauge transformation $B \mapsto B + d \Lambda$,
the Chan-Paton gauge fields necessarily undergo an affine
translation $A \mapsto A - \Lambda$.  As a result, since orbifold group
actions on $B$ fields in general mix group actions on the base space
with gauge transformations of the $B$ field, they also act nontrivially
on Chan-Paton factors, and hence on D-branes.

In \cite{medt} we described orbifold group actions on D-branes
for the most general possible orbifold group action on the $B$ field.
Assuming for simplicity that the $B$ field is topologically trivial,
we can write the most general orbifold group action as
\begin{eqnarray*}
g^* A^{\alpha} & = & \left( \gamma^g_{\alpha} \right) \,
A^{\alpha} \, \left( \gamma^g_{\alpha} \right)^{-1} \: + \:
\left( \gamma^g_{\alpha} \right) \,
d \left( \gamma^g_{\alpha} \right)^{-1} \: + \:
I \Lambda(g)^{\alpha} \\
g^* g_{\alpha \beta} & = & \left( T^g_{\alpha \beta} \right) \,
\left[ \, \left( \gamma^g_{\alpha} \right) \,
\left( g_{\alpha \beta} \right) \,
\left( \gamma^g_{\beta} \right)^{-1} \, \right] \\
\left( \omega^{g_1, g_2}_{\alpha} \right) \,
\left( \gamma^{g_1 g_2}_{\alpha} \right) & = &
\left( g_2^* \gamma^{g_1}_{\alpha} \right) \,
\left( \gamma^{g_2}_{\alpha} \right)
\end{eqnarray*}
where $\gamma^g_{\alpha}$ defines an action on the Chan-Paton factors,
and $I$ denotes the identity matrix.

In the case of ordinary discrete torsion, recall again that the
bundles $T^g$ were trivial, the connections $\Lambda(g)$ gauge-trivial,
and the connection-preserving bundle maps $\omega^{g_1,g_2}$ become
constants defining the group cocycles.  In this case, the expression
above simplifies to become
\begin{eqnarray*}
g^* A^{\alpha} & = & \left( \gamma^g_{\alpha} \right) \,
A^{\alpha} \, \left( \gamma^g_{\alpha} \right)^{-1} \: + \:
\left( \gamma^g_{\alpha} \right) \,
d \left( \gamma^g_{\alpha} \right)^{-1} \\
g^* g_{\alpha \beta} & = & 
\left[ \, \left( \gamma^g_{\alpha} \right) \,
\left( g_{\alpha \beta} \right) \,
\left( \gamma^g_{\beta} \right)^{-1} \, \right] \\
\left( \omega^{g_1, g_2}_{\alpha} \right) \,
\left( \gamma^{g_1 g_2} \right) & = &
\left( g_2^* \gamma^{g_1}_{\alpha} \right) \,
\left( \gamma^{g_2}_{\alpha} \right)
\end{eqnarray*}
which just describes a projectivized orbifold group action on the D-brane
worldvolume, precisely as predicted in \cite{doug1,doug2}.

Next, consider the case of toroidal orbifolds and our non-discrete-torsion
degrees of freedom.  Here, again the bundles $T^g$ are trivial,
as all flat bundles on a torus are trivial, and we have turned off
discrete torsion so that all the $\omega^{g_1,g_2}$ are trivial,
but the $\Lambda(g)$ are still nontrivial, and so we obtain
\begin{eqnarray*}
g^* A^{\alpha} & = & \left( \gamma^g_{\alpha} \right) \,
A^{\alpha} \, \left( \gamma^g_{\alpha} \right)^{-1} \: + \:
\left( \gamma^g_{\alpha} \right) \,
d \left( \gamma^g_{\alpha} \right)^{-1} \: + \:
I \Lambda(g)^{\alpha} \\
g^* g_{\alpha \beta} & = & 
\left( T^g_{\alpha \beta} \right) \,
\left[ \, \left( \gamma^g_{\alpha} \right) \,
\left( g_{\alpha \beta} \right) \,
\left( \gamma^g_{\beta} \right)^{-1} \, \right] \\
\left( \gamma^{g_1 g_2}_{\alpha} \right) & = &
\left( g_2^* \gamma^{g_1}_{\alpha} \right) \,
\left( \gamma^{g_2}_{\alpha} \right)
\end{eqnarray*}
Put another way, under the action of the orbifold group,
instead of projectivizing the orbifold group action as in
\cite{doug1,doug2}, instead the orbifold group maps 
the bundle ${\cal E} \mapsto {\cal E} \otimes T^g$.
Because the bundles $T^g$ obey the group law in the appropriate sense,
this orbifold group action is an honest orbifold group action, and not
a projectivized action.

Note in hindsight that our result can be derived much more simply
(and much more loosely) from the general idea that 
gauge transformations that rotate the
period of the $B$ field by $2 \pi$ are equivalent to tensoring the
D-brane bundles with some line bundle.
Thus, the fact that we are seeing these orbifold degrees of freedom
involving nonzero gauge fields $\Lambda(g)$'s correspond to 
tensoring D-brane bundles should not be a surprise.

In passing, note that we are {\it not} merely making an ansatz
that D-brane actions have such a form; rather, we are recovering
these D-brane actions as a special case of the general first-principles
derivation
presented in \cite{medt}.

\section{First consistency check:  Douglas's calculation} \label{consistency1}

In \cite{doug1,pauldt}, a calculation was performed
to check the consistency of the D-brane orbifold group
realization of discrete torsion \cite{doug1}, with
the usual closed-string twisted sector phases of \cite{vafa1}.

In this calculation, one takes a Riemann surface with boundary,
and evaluates the Wilson loop along the boundary.  The evaluation of
that Wilson loop will pick up a phase, as determined by the orbifold
group action on the D-branes, and for consistency, the total phase
one gets should be the same as for the corresponding closed-string
twisted sector.

\begin{figure}
\centerline{\psfig{file=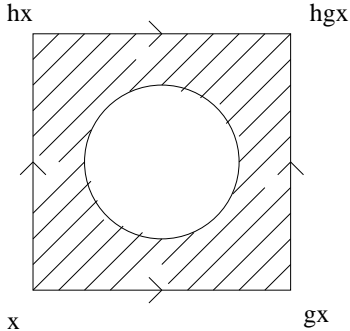,width=2in}}
\caption{\label{oneloopfig} Genus one Riemann surface with boundary}
\end{figure} 

An example will make this calculation much more clear.
Consider a genus one Riemann surface with boundary, describing
a twisted sector, as illustrated in figure~\ref{oneloopfig}.
Let us evaluate the Wilson loop around the open string boundary,
in a limit where the boundary has been moved close to the edges of
the square.  We should find that the orbifold group action on D-branes
modifies the Wilson loop calculation so as to give a phase factor,
which should be the same phase factor as for the corresponding
closed-string one-loop twisted sector.

To illustrate the method, we shall at first review this calculation
for ordinary discrete torsion, then we shall check that the same method
applied to our momentum shift degrees of freedom does correctly
match our D-brane action, described above, with closed string
twisted sector phase factors described previously.

The Wilson loop around the boundary shown in figure~\ref{oneloopfig}
has the form
\begin{equation}
\mbox{Tr }
\left( P \exp \int_x^{gx} A dx \right) \,
\left( P \exp \int_{gx}^{hgx} A dx \right) \,
\left( P \exp \int_{hgx}^{hx} A dx \right) \,
\left( P \exp \int_{hx}^x A dx \right)
\end{equation}
which we can rewrite as
\begin{equation} \label{simpdougloop}
\mbox{Tr }
\left( P \exp \int_x^{gx} A dx \right) \,
\left( P \exp \int_{x}^{hx} (g^* A) dx \right) \,
\left( P \exp \int_{gx}^{x} (h^* A) dx \right) \,
\left( P \exp \int_{hx}^x A dx \right)
\end{equation}
Using the relationship
\begin{displaymath}
g^* A \: = \: \left( \gamma^g \right) \,
A^{\alpha} \, \left( \gamma^g \right)^{-1} \: + \:
\left( \gamma^g \right) \,
d \left( \gamma^g \right)^{-1}
\end{displaymath}
discussed in the previous section, we can write
\begin{eqnarray*}
P \exp \left( \int_x^{hx} (g^* A) dx \right) & = &
\left( \gamma^g_x \right) \,
\left( P \exp \int_x^{hx} A dx \right) \,
\left( \gamma^g_{hx} \right)^{-1} \\
P \exp \left( \int_{gx}^x (h^* A) dx \right) & = &
\left( \gamma^h_{gx} \right) \,
\left( P \exp \int_{gx}^x A dx \right) \,
\left( \gamma^h_x \right)^{-1}
\end{eqnarray*}
Using the fact that $\gamma^g_{hx} = h^* \gamma^g_x$
and the relationship
\begin{displaymath}
\left( \omega^{g_1, g_2} \right) \gamma^{g_1 g_2} \: = \:
\left( g_2^* \gamma^{g_1} \right) \, \left( \gamma^{g_2} \right)
\end{displaymath}
we can simplify the Wilson loop~(\ref{simpdougloop})
to the form
\begin{eqnarray}
\lefteqn{
\left( \frac{ \omega^{h,g} }{ \omega^{g,h} } \right) \:
\mbox{Tr }  \left\{ 
\left( P \exp \int_x^{gx} A dx \right) \,
\left( \gamma^g_x \right) \,
\left( P \exp \int_x^{hx} A dx \right) \,
\left( \gamma^h_x \right) \, \right. }  \nonumber \\
& & \cdot \, \left.
\left( \gamma^g_x \right)^{-1} \,
\left( P \exp \int_x^{gx} A dx \right)^{-1} \,
\left( \gamma^h_x \right)^{-1} \,
\left( P \exp \int_x^{hx} A dx \right)^{-1} \right\}  \label{dtdougloop}
\end{eqnarray}

In other words, by using the orbifold group action on D-branes
appropriate for discrete torsion, we have recovered the closed string
one-loop twisted sector phase factor for discrete torsion,
just as in \cite{doug1,pauldt}.

Next we shall apply the same method to check the consistency of
the momentum shift degrees of freedom described in this paper,
by computing their closed-string one-loop twisted sector phase factor
using the corresponding orbifold group action on D-branes.

Again, consider the Riemann surface with boundary illustrated in
figure~\ref{oneloopfig}, and again, let us calculate the Wilson loop
around the boundary.  That Wilson loop again has the form~(\ref{simpdougloop}).
So, all that remains to do is to simplify that expression using 
the orbifold group action on D-branes for the momentum shift degrees
of freedom discussed in this paper.

Specifically, we shall use the relation
\begin{displaymath}
g^* A \: = \: \left( \gamma^g \right) \,
A^{\alpha} \, \left( \gamma^g \right)^{-1} \: + \:
\left( \gamma^g \right) \,
d \left( \gamma^g \right)^{-1} \: + \:
I \Lambda(g) 
\end{displaymath}
discussed in the previous section.
Using that relation, we find that 
\begin{eqnarray*}
P \exp \left( \int_x^{hx} (g^* A) dx \right) & = &
\left( \gamma^g_x \right) \,
\left( P \exp \int_x^{hx} A dx \right) \,
\left( \gamma^g_{hx} \right)^{-1}
\left( \exp \int_x^{hx} \Lambda(g)  \right) \\
P \exp \left( \int_{gx}^x (h^* A) dx \right) & = &
\left( \gamma^h_{gx} \right) \,
\left( P \exp \int_x^{gx} A dx \right)^{-1} \,
\left( \gamma^h_x \right)^{-1} \,
\left( \exp - \int_x^{gx} \Lambda(h)  \right)
\end{eqnarray*}
Working through the same algebra as above, it is easy to check
that the Wilson loop computation now simplifies to the same
form as~(\ref{dtdougloop}), 
except that the phase factor $\omega^{h,g}/\omega^{g,h}$
is replaced by the phase factor
\begin{displaymath}
\exp \left( \, \int_x^{hx} \Lambda(g)  \: - \:
\int_x^{gx} \Lambda(h)  \, \right)
\end{displaymath}
which the reader will recognize as the closed-string twisted sector
phase factor for these momentum shift degrees of freedom,
the same phase factor that simplifies to the form
$\exp i ( p_L a_R - p_R a_L)$ for toroidal orbifolds.

In other words, we have successfully checked, using the methods
of \cite{doug1,pauldt} that the orbifold group action on D-branes
that we have described for these momentum shift degrees of freedom
is indeed consistent with the corresponding closed-string description.

\section{Examples }
\label{vwanalogue}

\subsection{ $T^4/{\bf Z}_2$ spectrum computation}

To help illustrate these degrees of freedom, we will consider
the example of $T^4/{\bf Z}_2$.
In this example, combining the orbifold group action with
momentum shifts will have no effect -- the new orbifold CFT
is isomorphic to the original orbifold CFT \cite{katrinpriv}.
However, this example will provide us with the opportunity to explore
spectrum computations in some detail.

Note that it is not possible to turn on ordinary
discrete torsion in this orbifold: 
\begin{displaymath}
H^2({\bf Z}_2, U(1)) = 0.
\end{displaymath}
However, our momentum shift degrees of freedom are not counted by
$H^2(G,U(1))$, and they {\it can} be turned on in $T^4/{\bf Z}_2$.

To specify these momentum shift degrees of freedom, we must specify
a flat $U(1)$ gauge field on the covering space $T^4$
for each element of the orbifold group,
such that the gauge fields obey the group law.
Now, we can specify such gauge fields by their holonomies around
four cycles on $T^4$, and for those holonomies to obey the group
law simply means that their square must be trivial.
Thus, the possible holonomies around each circle are zero and
$1/2$ (in conventions where the circles have circumference 1).

In passing, note that since we are discussing {\it symmetric} orbifolds,
level-matching should be satisfied automatically.  We will not discuss
level-matching further.

Now, we would like to understand the orbifold group action on the
states of the theory.  
In a symmetric orbifold, the possible phases $\exp(i p \cdot a)$ on
a given state are restricted to only some $a_{L,R}$.  Depending upon
sign conventions, for $B \equiv 0$ in a symmetric orbifold the
possible phase factors arising from the group action on a single
state have either $a_L = a_R$ or $a_L = -a_R$ \cite{dgh,katrin}.
The one-loop twisted sector phase factors can have more general
$a_{L,R}$, even in a symmetric orbifold, but the phase factors arising
from the group action on a particular state are less general.

Let us compare to our general analysis, and describe such phase
factors in terms of integrals of $\Lambda(g)$'s.
If we work in a Hamiltonian formulation,
then following \cite{vafaed} it is natural to conjecture that under the
action of $h \in G$, a given state in the $g$th twisted sector should
be multiplied by the phase factor
\begin{displaymath}
\exp \left( \int_x^{gx} \Lambda(h)  \right)
\end{displaymath}
integrated along the length of the closed string.
(Note that the twisted sector state implicitly specifies the
path over which this integral should be performed, and that
since $\Lambda(h)$ is a flat connection, the integral only depends
upon topological characteristics of the path.)
It is straightforward to check that this is the same as the phase
factor $\exp(i p \cdot a)$ for $a_L = a_R$, precisely right for
a symmetric orbifold (modulo sign conventions) \cite{dgh,katrin}.
It is also straightforward to check, following \cite{vafaed},
that this phase factor obeys
the group law, using the relation \cite{medt}
\begin{displaymath}
\Lambda(g_1 g_2) \: = \: \Lambda(g_2) \: + \: g_2^* \Lambda(g_1) \: - \:
d \log \omega^{g_1,g_2}
\end{displaymath}
In the present case, since we have assumed no discrete torsion,
the $\omega$ are all trivial.
Acting with $h_1$ and then $h_2$, from the analysis above,
would give a phase factor
\begin{displaymath}
\exp\left( \int_x^{gx} \Lambda(h_2)  \right) \,
\exp\left( \int_x^{gx} h_2^* \Lambda(h_1)  \right)
\end{displaymath}
but an immediate consequence of the relation above is that this
phase factor is the same as
\begin{displaymath}
\exp \left( \int_x^{gx} \Lambda(h_1 h_2)  \right)
\end{displaymath}
hence the group law is obeyed.
As a further check, in the next section, 
we shall confirm this description of the group
action on states by repeating Gomis's consistency check \cite{gomis}
between group actions on closed string twisted sector states,
and group actions on D-branes.

Now, the reader might be slightly confused by the relationship
between the phase factor postulated above and the corresponding
phase factor for discrete torsion described in \cite{vafaed}.
There, a state in the $g$th twisted sector picked up the same
phase factor under the action of $h \in G$ as the one-loop 
$(g,h)$ twisted
sector phase factor, which had the effect of insuring that inserting
$\sum_G h$ inside correlation functions duplicated $(g,h)$ twisted sectors.
Here, by contrast, the $h$ action on a $g$ twisted sector state
does {\it not} give the complete phase factor, only part of it.
Again, this group action on states has appeared previously
in \cite{dgh}, so it should be correct, though the comparison
to the discussion in \cite{vafaed} might confuse the reader.
Although we have not checked thoroughly, we believe the rest
of the $T^2$ phase factor comes from the group action on the Hamiltonian
describing time-evolution of the states.
Recall from expression~(\ref{phstart}) that 
the one-loop twisted sector phase factor has the form
\begin{displaymath}
\exp\left( \int_x^{gx} \Lambda(h) \: - \: \int_x^{hx} \Lambda(g) \right)
\end{displaymath}
The first integral can be interpreted as the integral along the closed
string in a $g$ twisted sector, {\it i.e.} an integral along the
spacelike directions, but the second integral must be
interpreted as an integral along the timelike directions, which cannot
possibly come out of a group action on a state at a fixed time.
Rather, the second integral must necessarily emerge
from a nontrivial group action on the Hamiltonian, which {\it is}
integrated along the timelike direction.

In the case of $T^4/{\bf Z}_2$, the effect of this group action
on massless states is relatively minimal.  For generic tori,
all the massless states correspond to strings of length zero -- this
is clear in untwisted sectors, and in twisted sectors, the vacua 
all correspond to strings sitting at the isolated fixed points.
If the strings all have length zero, then our phase factor
\begin{displaymath}
\exp\left( \int_x^{gx} \Lambda(h)  \right)
\end{displaymath}
is necessarily trivial in each case, and so the massless
spectrum is unaffected.

Now, for nongeneric tori, by arranging for some of the radii
to have special values, one can get enhanced gauge symmetries
in the ordinary way.  

Let us consider specifically bosonic $T^4/{\bf Z}_2$, for simplicity,
and we shall follow the conventions of \cite[section 10.1]{lt}
for orthogonal tori and vanishing $B$ field (for simplicity).
For the $i$th circle, in the untwisted sector we can write
\begin{eqnarray*}
p_L^i & = & \frac{M_i}{R_i} \: + \: \frac{1}{2}L_i R_i \\
p_R^i & = & \frac{M_i}{R_i} \: - \: \frac{1}{2}L_i R_i 
\end{eqnarray*}
where $R_i$ is the radius of the circle, $M_i$ is the momentum,
and $L_i$ is the winding number; $M_i, L_i \in {\bf Z}$.
(Recall that for expositional simplicity we are assuming $B \equiv 0$ 
in the first part of this paper; if $B$ were nonzero, there would
be additional terms in the expressions above for the momenta.)
The ${\bf Z}_2$ action maps $X \mapsto -X$, hence it acts on this
momentum lattice as $p_L,R^i \mapsto - p_L,R^i$.
For special radii, it is well-known that one gets enhanced gauge
symmetries.  For example, for a single circle, 
using $|M,L>$ to denote a soliton vacuum with momentum $M$ and
winding number $L$,
there are spacetime vectors $\alpha^{\mu}_{-1}| \pm 1, \pm 1>$
(where where $\alpha^{\mu}_n$ are the oscillator mode expansion terms
of $\partial X^{\mu}$),
that become massless when $R=1/\sqrt{2}$.

Without turning on the shift orbifold phases discussed in this paper,
linear combinations such as
\begin{displaymath}
\alpha^{\mu}_{-1}\left( | +1,+1> \: + \: |-1,-1> \right)
\end{displaymath}
(corresponding to low-energy vector fields),
survive the orbifold projection, and for the correct radius,
will give massless vectors, {\it i.e.}, enhanced gauge symmetry
at special points in moduli space.

How do our degree of freedom act on such states?
Recall a given state in the $g$th twisted sector
will pick up a phase factor $\exp\left( \int_x^{gx}\Lambda(h) 
\right)$. 
Now, in our conventions, that integral is performed over $\sigma$ directions,
so for an untwisted sector state, we can rewrite that phase factor as
\begin{displaymath}
\exp\left( \int \Lambda_i \frac{\partial X^i}{\partial \sigma} d \sigma \right)
\end{displaymath}
In our conventions, 
\begin{displaymath}
\frac{ \partial X^i }{ \partial \sigma } \: = \: p_L^i \: - \: p_R^i
\: = \: L_i R_i
\end{displaymath}
and the integration is performed along the closed string.
A nontrivial holonomy along the $i$th circle, obeying
the ${\bf Z}_2$ group law, would be described by
$\Lambda_i = 2 \pi i (\frac{1}{2}) / R_i$,
so in particular, for $L_i = 1$, we compute that the phase factor is
given by
\begin{displaymath}
\exp\left( \int \Lambda dX \right) \: = \: -1
\end{displaymath}
(at least, for a circle with winding number one and nonzero holonomy $\Lambda$).

In particular, soliton states that previously survived the
orbifold group projection, {\it e.g.}, low-energy vector fields
\begin{displaymath}
\alpha^{\mu}_{-1}\left( | +1,+1> \: + \: |-1,-1> \right)
\end{displaymath}
now pick up a sign under the group action, and so are no longer invariant.
On the other hand, some states that were previously projected out
of the spectrum, such as the low-energy vector fields
\begin{displaymath}
\alpha^{\mu}_{-1}\left( | +1,+1> \: - \: |-1,-1> \right)
\end{displaymath}
are now ${\bf Z}_2$-invariant and so
retained in the spectrum.  The net effect is not to change
the spectrum of the low-energy effective theory
\cite{katrinpriv}, though the vertex operators are permuted as we
see here.

Thus, we see that one result of turning on one of our degrees of freedom
is to play with some of the solitonic states responsible for enhanced
gauge symmetries at special points in the toroidal moduli space.

\subsection{ $T^2/{\bf Z}_2$ spectrum computation}

This example will not be supersymmetric, but the orbifold theory
produced by turning on momentum shift phases will differ from the
original orbifold, unlike our last example.

Consider the action of ${\bf Z}_2$ on $T^2$ that maps
\begin{displaymath}
\left( x_1, x_2 \right) \: \mapsto \: \left( x_1, - x_2 \right)
\end{displaymath}
{\it i.e.} the ${\bf Z}_2$ action on the base space only acts on the
second $S^1$.  We will combine this ${\bf Z}_2$ action on the base
with a momentum shift phase factor that only shifts along the $x_1$
direction.  (This orbifold has been described previously as, {\it e.g.},
lattice 10 in \cite{katrinc=2}.)

If we let $g$ denote the nontrivial element of ${\bf Z}_2$,
then we can describe those shift phases by taking 
\begin{displaymath}
\Lambda(g) \: = \: \frac{1}{2} dx_1
\end{displaymath}
where, in poor notation, we have used ``$dx_1$'' to denote the
closed non-exact generator of cohomology of the first $S^1$.
The fact that there is no shift along the second circle
is reflected in the fact that $\Lambda(g)$ has no holonomy along the
second circle.

We shall not perform a full spectrum calculation, but instead will
merely outline some results.
If we let
\begin{displaymath}
| M_1, L_1; M_2, L_2 >
\end{displaymath}
denote a soliton state on the $T^2$, with momentum and winding along
the two circles denoted by $M_{1,2}$, $L_{1,2}$,
then the analysis proceeds much as for $T^4/{\bf Z}_2$.
Following the same general methods as in $T^4/{\bf Z}_2$,
it is straightforward to check that untwisted-sector states such as
\begin{displaymath}
\alpha^2_{-1} | +1, +1; 0,0 >
\end{displaymath}
are invariant under the orbifold group action.
In this example, $\alpha_{-1}^2$ is the $-1$ mode in the Fourier expansion
of $x_2(\sigma, \tau)$, and so is multiplied by $-1$ by the group action.
The soliton state $| +1, +1; 0,0>$ is invariant under the group action
on the base space $T^2$, since it has no winding around the second $S^1$,
but because it has nontrivial winding around
the first $S^1$, it picks up a momentum shift phase factor,
which is easily computed to be $-1$.  Hence, this state is invariant
under the group action and survives to describe a low-energy scalar.

Similarly, the (untwisted sector) low-energy vector field described by
\begin{displaymath}
\alpha^{\mu}_{-1} | 2,0; 0,0>
\end{displaymath}
also survives the orbifold projection.
This follows because $\alpha^{\mu}$, where $\mu$ corresponds to
directions orthogonal to the $T^2$, is invariant under the ${\bf Z}_2$
action, and the soliton state $|2,0;0,0>$ also picks up no net phase
from the momentum shift phase factor.

It is also straightforward to check that the (untwisted sector) state
\begin{displaymath}
\alpha^{\mu}_{-1} \left( | 0,0; +1,+1> + |0,0;-1,-1> \right)
\end{displaymath}
is invariant under the group action.
Here, $\alpha^{\mu}$ is invariant as above.
The ${\bf Z}_2$ action on the base space exchanges 
$|0,0;\pm 1, \pm 1>$, and since there is no winding in the
first $S^1$, there is no phase factor due to the momentum lattice
shift factor.

\section{Second consistency check:  Gomis's calculation} \label{consistency2}

In \cite{gomis} a consistency check was performed on the
ansatz of \cite{doug1,doug2}.  In particular, in \cite{gomis}
consistency of open-closed string interactions was used to
argue that, given the closed-string twisted sector phase factor
of \cite{vafa1}, the ansatz of \cite{doug1,doug2} for orbifold
group actions on D-branes was a necessary consequence.

\begin{figure}
\centerline{\psfig{file=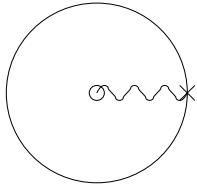,width=1in}}
\caption{\label{gcalc} Disk amplitude with a twisted sector state
inserted in the center of the disk.  A branch cut runs from the center
to the edge.}
\end{figure}

The calculation of \cite{gomis} involved a disk amplitude
with a Ramond-Ramond closed-string state in the $g$-th twisted
sector, inserted in the center of the disk, and also a photon arising
from the open string ending on a D-brane transverse to the orbifold,
as schematically illustrated in figure~\ref{gcalc}.

Roughly following the notation of \cite{gomis}, 
we can schematically write the amplitude
in question as
\begin{equation}  \label{basicamp}
\mbox{Tr }\left\{ \gamma^g_x P \exp\left( \int_x^{g x} A dx \right) \right\} \,
< V^g(0) V_{\mu}(1) >
\end{equation}
where $V^g(0)$ is the vertex operator for the twist field,
$V_{\mu}(1)$ is the vertex operator inserted on the boundary,
and we have explicitly included the Wilson loop computation along the
open string boundary, for use later.
We must demand that this amplitude be $G$-invariant.

Under the action of $h \in G$, the amplitude above becomes
\begin{equation}  \label{hamp}
\mbox{Tr }\left\{ (h^* \gamma^g_x) P \exp\left( \int_x^{g x} (h^*A) dx \right) \right\} \,
< h \cdot V^g(0) V_{\mu}(1) >
\end{equation}
Let us evaluate the expression above both for ordinary discrete torsion,
to recover the results of \cite{gomis}, and for our momentum shift degrees
of freedom.

First, consider ordinary discrete torsion.
As described in detail in \cite{medt} and reviewed above,
\begin{eqnarray}
g^* A & = & \gamma^g A \left( \gamma^g \right)^{-1} \: + \:
( \gamma^g ) d ( \gamma^g )^{-1} \\
\left( \omega^{g_1, g_2} \right) \gamma^{g_1 g_2} & = &
\left( g_2^* \gamma^{g_1} \right) \left( \gamma^{g_2 } \right) \label{gammadtreln}
\end{eqnarray}
from which we see that
\begin{displaymath}
P \exp\left( \int_x^{g x} (h^*A) dx \right) \: = \:
\gamma^h_x \left( P \exp \int_x^{gx} A dx \right) \left( \gamma^h_{gx} \right)^{-1}
\end{displaymath}
so using the fact that $\gamma^h_{gx} \: = \: g^* \gamma^h_x$
and equation~(\ref{gammadtreln}) above, it is easy to check that
\begin{displaymath}
\mbox{Tr } \left\{ (h^*  \gamma^g_x) P \exp\left( \int_x^{g x} (h^*A) dx \right) 
\right\} \: = \: \left( \frac{ \omega^{g,h} }{ \omega^{h,g} } \right)
\mbox{Tr }\left\{ \gamma^g_x P \exp\left( \int_x^{g x} A dx \right) \right\}
\end{displaymath}

Finally, to show that ordinary discrete torsion leaves the
amplitude~(\ref{basicamp}) invariant, we use the fact that the
group action on a twisted sector state is nontrivial.
Following \cite{vafaed}, in our conventions
\begin{displaymath}
h \cdot V^g(0) \: = \: \left( \frac{ \omega^{h,g} }{ \omega^{g,h} } \right) 
V^g(0)
\end{displaymath}
hence equations~(\ref{hamp}) and~(\ref{basicamp}) are identical,
{\it i.e.}, the amplitude~(\ref{basicamp}) is $G$-invariant.

So far we have merely duplicated the results of \cite{gomis}.
Next, we shall perform the same calculation for the momentum shift degrees
of freedom that are the focus of this paper.

For these momentum shift degrees of freedom, we use the relation
\begin{displaymath}
g^* A \: = \: \left( \gamma^g \right) \,
A \, \left( \gamma^g \right)^{-1} \: + \:
\left( \gamma^g \right) \,
d \left( \gamma^g \right)^{-1} \: + \:
I \Lambda(g) 
\end{displaymath}
from which it is easy to see that
\begin{displaymath}
P \exp \left( \int_x^{gx} (h^* A) dx \right) \: = \:
\gamma^h_x \left( P \exp \int_x^{gx} A dx \right) \left( \gamma^h_{gx} \right)^{-1}
\left( \exp \int_x^{gx} \Lambda(h)  \right)
\end{displaymath}
Thus, we can immediately simplify part of the amplitude~(\ref{hamp})
to the form
\begin{displaymath}
\mbox{Tr } \left\{ (h^*  \gamma^g_x) P \exp\left( \int_x^{g x} (h^*A) dx \right) 
\right\} \: = \: \mbox{Tr } \left\{ \gamma^g_x P \exp \left( \int_x^{gx} 
A dx \right) \right\} \,
\left( \exp \int_x^{gx} \Lambda(h)  \right)
\end{displaymath}

Now, as we discussed in the previous section, a closed-string twisted
sector operator $V^g$ should pick up a factor of $\exp\left( \int_x^{gx}
\Lambda(h) \right)$ under the action of $h \in G$.  Thus, again,
we see that the amplitude~(\ref{basicamp}) is invariant under the orbifold
group, for these momentum shift degrees of freedom.
Thus, our momentum shift degrees of freedom pass Gomis's consistency test.

\section{M theory dual of IIA discrete torsion}  \label{mdual}

\subsection{Basic analysis}

One natural question one can ask, that to our knowledge has not
been previously answered in the literature, is simply,
what is the strong-coupling limit of discrete torsion?
Given IIA string theory compactified on an orbifold with
discrete torsion turned on, what is the corresponding M theory
limit of the theory?

Strictly speaking, there is another question one should ask first,
namely:  what is the M theory version of a string orbifold?
String orbifolds are defined in a manner that appears intrinsic
to CFT, and M theory has no such CFT-based description.
There has been some work (see {\it e.g.} \cite{ovrut} and references
therein) on defining orbifolds in M theory by making ansatzes
for matter content localized on fixed points, consistent with
anomaly cancellation.
If pressed for a more nearly first-principles construction,
we would observe the natural candidate for defining an orbifold
in M theory is as an M theory compactification on a quotient stack
\cite{meqs,kps}.
In any event,
we shall simply assume that orbifolds are sensibly-defined in M theory,
and will not discuss this technical point
further.

Given the assumption that orbifolds are well-defined in M theory,
how does IIA discrete torsion appear?  We have argued extensively
(see \cite{dtrev} for a review) that discrete torsion is a feature
of the $B$ field -- discrete torsion is an inevitable consequence
of defining orbifold group actions on any theory containing a $B$ field,
and is not in any sense specific to CFT.
However, M theory has no $B$ field.  Instead of a two-form potential $B$,
M theory has a three-form potential $C$, which descends to the IIA $B$ field
after `compactifying' the $C$ field on the eleven-dimensional circle.

Since M theory has no $B$ field, how can IIA discrete torsion, or IIA
shift orbifolds, manifest themselves?

The answer lies in an analogue for $C$ fields of the momentum shift
degrees of freedom
we have discussed elsewhere in this paper, which will turn out to
encode both IIA discrete torsion as well as IIA shift orbifolds.

We should add, the analysis we shall present in this subsection is closely
related to work in \cite{seki}.  Our purpose in this paper is
to relate certain obscure orbifold group actions to shift orbifolds,
and the M theory lift of discrete torsion provides an analogue
for the $C$ field.

As discussed in \cite{cdt}, orbifold group actions on $C$ fields have
degrees of freedom that are closely analogous to discrete torsion.
For example, one set of degrees of freedom is measured by
group cohomology, namely $H^3(G, U(1))$ (degree three instead of degree two).
These particular degrees of freedom\footnote{In the derivation \cite{cdt} of
these degrees of freedom, we ignored the flux quantization condition
\cite{edflux} on the M theory $C$ field.  This quantization condition
is believed \cite{triples} to restrict allowed elements of $H^3(G, U(1))$,
in much the same way that level-matching restricts possible 
asymmetric orbifolds.  We hope to publish a thorough study of
the effect of the flux-quantization condition on the $C$ field's
$H^3(G, U(1))$-enumerated degrees of freedom
in the future.  } manifest themselves in a manner
that is very similar to discrete torsion -- a `twisted sector' membrane
wrapped on $T^3$ will pick up a phase of the form
\begin{equation}    \label{t3phasebasic}
\left( \omega^{g_1, g_2, g_3} \right) \,
\left( \omega^{g_2, g_1, g_3} \right)^{-1} \,
\left( \omega^{g_3, g_2, g_1} \right)^{-1} \,
\left( \omega^{g_3, g_1, g_2} \right) \,
\left( \omega^{g_2, g_3, g_1} \right) \,
\left( \omega^{g_1, g_3, g_2} \right)^{-1}
\end{equation}
where $\omega$ is a group 3-cocycle, and $g_1$, $g_2$, $g_3$ are
three commuting elements of $G$, defining the membrane twisted sector.
This expression, although written in terms of cocycles, descends
to a well-defined function on group cohomology, and moreover possesses
an $SL(3,{\bf Z})$-invariance that is closely analogous to modular invariance
for closed strings.

However, these are not the only degrees of freedom appearing in
orbifold group actions on $C$ fields, just as discrete torsion is not the
only degree of freedom appearing in orbifold group actions on $B$ fields.
As described in \cite{cdt}, the difference between any
two orbifold group actions on a $C$ field is defined
by a set of 1-gerbes with flat connection $({\cal B}(g)^{\alpha},
{\cal A}(g)^{\alpha \beta}, \Upsilon^g_{\alpha \beta \gamma})$,
one for each $g \in G$, plus maps $\left(\Omega^{g_1, g_2},
\theta(g_1, g_2) \right)$ between 1-gerbes with connection
\begin{displaymath}
\left( {\cal B}(g_2)^{\alpha}, {\cal A}(g_2)^{\alpha \beta},
\Upsilon^{g_2}_{\alpha \beta \gamma} \right)
\otimes g_2^* \left(
{\cal B}(g_1)^{\alpha}, {\cal A}(g_1)^{\alpha \beta},
\Upsilon^{g_1}_{\alpha \beta \gamma} \right) \: \longrightarrow \:
\left( {\cal B}(g_1 g_2)^{\alpha}, {\cal A}(g_1 g_2)^{\alpha \beta},
\Upsilon^{g_1 g_2}_{\alpha \beta \gamma} \right)
\end{displaymath}
and ``maps between maps'' $\omega^{g_1, g_2, g_3}$:
\begin{displaymath}
\omega^{g_1, g_2, g_3}: \: \Omega^{g_1, g_2 g_3} \circ \Omega^{g_2, g_3}
\: \longrightarrow \: \Omega^{g_1 g_2, g_3} \circ
g_3^* \Omega^{g_1, g_2}
\end{displaymath}
The 1-gerbe morphisms $\left(
\Omega^{g_1, g_2}, \theta(g_1, g_2) \right)$ are constrained
to make the following diagram commute
\begin{equation}   \label{upcocycle}
\xymatrix@C+40pt{
\Upsilon^{g_3} \otimes g_3^* \left( \Upsilon^{g_2} \otimes
g_2^* \Upsilon^{g_1} \right)
\ar[r]^{g_3^* \Omega^{g_1, g_2}}
\ar[d]_{\Omega^{g_2, g_3}} 
& \Upsilon^{g_3} \otimes g_3^* \Upsilon^{g_1 g_2}
\ar[d]^{\Omega^{g_1 g_2, g_3}} \\
\Upsilon^{g_2 g_3} \otimes (g_2 g_3)^* \Upsilon^{g_1}
\ar[r]^{ \Omega^{g_1, g_2 g_3} }
& \Upsilon^{g_1 g_2 g_3}
}
\end{equation}
up to isomorphisms of maps defined by the $\omega^{g_1, g_2, g_3}$ above,
and the maps of maps $\omega^{g_1, g_2, g_3}$ are constrained to obey
\begin{displaymath}
\omega^{g_1 g_2, g_3, g_4} \circ \omega^{g_1, g_2, g_3 g_4} \: = \:
g_4^* \omega^{g_1, g_2, g_3} \circ \omega^{g_1, g_2 g_3, g_4} \circ
\omega^{g_2, g_3, g_4}
\end{displaymath}
Moreover, the data above obeys two levels of gauge invariances:
gauge transformations of the 1-gerbes, and gauge transformations
of the maps between the 1-gerbes.

Let us build a $C$ action from a $B$ action as follows.
Let $d x^{11}$ denote the generator of cohomology\footnote{Note our notation
is appalling, as ``$dx^{11}$'' here is closed but {\it not} exact.} 
of the eleven-dimensional
circle.  Recall that the IIA $B$ field is obtained from the
M theory $C$ field as, $C = B \wedge d x^{11}$, which suggests the procedure
to follow.  Gauge transformations of the $B$ field, which defined the
different IIA degrees of freedom, become gauge transformations of the
$C$ field, after wedging with $dx^{11}$, hence gauge fields become
$B$ fields, and so forth.  By taking the gauge transformations
defining the various orbifold group actions on the $B$ field,
and suitably wedging with $dx^{11}$, it is clear that one should
recover the M theory interpretation of IIA $B$ field orbifold group
actions.

Let us work through the details for both IIA discrete torsion,
as well as the other degrees of freedom discussed in this paper.

Recall ordinary discrete torsion, viewed as an orbifold group action
on the $B$ field, is defined by taking the bundles with connection
to be trivial, with gauge-trivial connection; discrete torsion is encoded
in the connection-preserving bundle isomorphisms that preserve the group law.
In terms of the $C$ field data above, if we take the 1-gerbes to be
trivial, with gauge-trivial connection, then the maps 
$\left(\Omega^{g_1, g_2},
\theta(g_1, g_2) \right)$ are equivalent to principal $U(1)$ bundles
$\Omega^{g_1, g_2}$ with connection $\theta(g_1, g_2)$.
Take the bundles $\Omega^{g_1,g_2}$ to be trivial, with flat
connection defined by $\left( \log \omega_{g_1,g_2} \right) d x^{11}$,
so that the holonomy around the eleven-dimensional circle is $\omega_{g_1,g_2}$,
where $\omega_{g_1,g_2}$ is the group 2-cocycle defining the relevant
element of IIA discrete torsion.  (The log is necessary
as the $\omega$'s are Lie group valued, yet the gauge field is Lie algebra
valued.)  With those definitions, it is straightforward
to check that diagram~(\ref{upcocycle}) commutes exactly, so we can take
the maps $\omega^{g_1,g_2,g_3}$
to all be trivial.
The remaining residual gauge transformations merely change the
group 2-cocycles $\omega_{g_1,g_2}$ by coboundaries.
Thus, we have duplicated IIA discrete torsion as an unusual action
on the M theory $C$ field.

The reader might well ask, what is the M theory dual of the other
IIA degrees of freedom discussed in this paper?
This question is also straightforward to answer.
Recall that for toroidal orbifolds, these other degrees of freedom
were described by trivial bundles $T^g$ with flat nonzero connections
$\Lambda(g)$, and for simplicity we took the isomorphisms
$\omega_{g_1,g_2}$ to contain no new information.
Define an orbifold group action on the $C$ field as follows.
Take the 1-gerbes $\Upsilon^g$ to all be trivial,
with flat connections defined by ${\cal B}(g) = \Lambda(g) \wedge d x^{11}$.
With these choices, we can take the maps $(\Omega^{g_1,g_2},
\theta(g_1,g_2))$ to be trivial, and so the $\omega^{g_1,g_2,g_3}$
are also trivial.
In other words, the flat bundles $T^g$ defining the IIA degrees of 
freedom, become flat 1-gerbes in M theory.

In passing, we should mention there has been some previous work on
understanding discrete torsion in the context of M theory lifts 
of orientifold planes \cite{kol1,kol2}, in the general
spirit of \cite{ovrut}.  We have nothing to say
about orientifolds in this paper; however, for completeness,
we feel their work deserves mention.

\subsection{Consistency check:  twisted sector phases from wrapped
membranes}

Although the statements above both seem straightforward,
we shall check both statements by computing phase factors from
membranes on $T^3$ twisted sectors.  In particular, if we wrap
a membrane on the eleven-dimensional circle, we should recover a IIA string.
In this manner, we will be able to explicitly recover the usual
IIA twisted sector phase factors from membrane phase factors due
to the $C$ field.

\begin{figure}
\centerline{\psfig{file=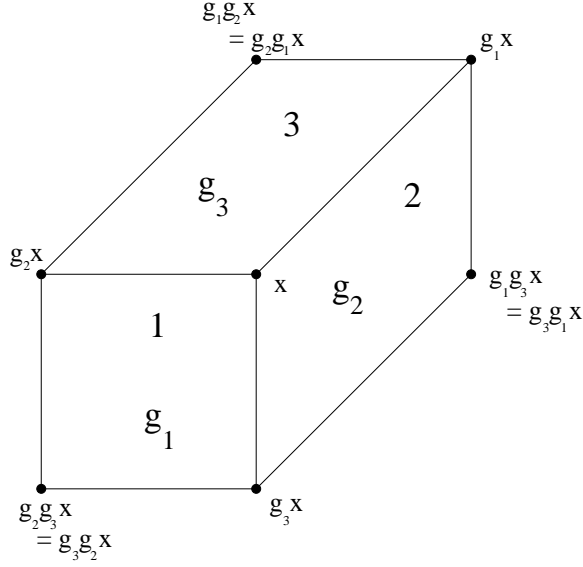,width=3in}}
\caption{ \label{t3phase} Membrane $T^3$ twisted sector}
\end{figure}

Consider a membrane $T^3$ twisted sector, as illustrated in 
figure~\ref{t3phase}.  This twisted sector is defined by three
commuting elements $g_1$, $g_2$, $g_3$ of the orbifold group.
For the purposes of deriving IIA twisted sector phases from M theory membranes,
we shall assume $g_3=1$, that the path $x \rightarrow g_3 \cdot x$ 
winds once around the eleven-dimensional circle, and that the other
paths $x \rightarrow g_1 \cdot x$, $x \rightarrow g_2 \cdot x$ do not
wind at all around the eleven-dimensional circle.

Next, the phase factor a membrane in such a twisted sector would
pick up from orbifold group actions on the $C$ field is given by
\begin{eqnarray}
\lefteqn{ \left( \omega^{g_1, g_2, g_3}_x \right) \,
\left( \omega^{g_2, g_1, g_3}_x \right)^{-1} \,
\left( \omega^{g_3, g_2, g_1}_x \right)^{-1} \,
\left( \omega^{g_3, g_1, g_2}_x \right) \,
\left( \omega^{g_2, g_3, g_1}_x \right) \,
\left( \omega^{g_1, g_3, g_2}_x \right)^{-1} }  \nonumber \\
\: & \: \: & \: \cdot \exp \, \left( \, - \: \int^{g_3 x}_x \left[ \, 
\theta(g_1, g_2) \: - \: \theta(g_2, g_1) \, \right] \: - \:
\int^{g_1 x}_x \, \left[ \, \theta(g_2, g_3) \: - \: \theta(g_3, g_2) \, 
\right] \, \right) \nonumber \\
\: & \: \: & \: \cdot \exp \, \left( 
\: - \:
\int^x_{g_2 x} \, \left[ \, \theta(g_1, g_3) \: - \: \theta(g_3, g_1) \, 
\right]
\, \right)  \nonumber \\
 \: & \: \: & \: \cdot \exp \, \left(
\, \int_1 {\cal B}(g_1) \: + \:
\int_2 {\cal B}(g_2) \: + \: \int_3 {\cal B}(g_3) \, \right)
\label{cfieldphase}
\end{eqnarray}
as derived in \cite{cdt}.

Consider first the M theory lift of IIA discrete torsion, as we described
it above.  Since the 1-gerbes are trivial, with gauge-trivial connections
${\cal B}(g)$, the integrals over ${\cal B}$'s are all trivial,
and similarly the $\omega^{g_1,g_2,g_3}$ factors are also trivial.
Taking $g_3=1$, so that the path $x \rightarrow g_3 \cdot x$ winds
once around the eleven-dimensional circle, the phase factor~(\ref{cfieldphase})
simplifies to become
\begin{eqnarray*}
\lefteqn{
\exp \, \left( \, - \: \int^{g_3 x}_x \left[ \, 
\theta(g_1, g_2) \: - \: \theta(g_2, g_1) \, \right] \: - \:
\int^{g_1 x}_x \, \left[ \, \theta(g_2, g_3) \: - \: \theta(g_3, g_2) \, 
\right] \, \right) } \\
& \: & \: \cdot \exp \, \left( 
\: - \:
\int^x_{g_2 x} \, \left[ \, \theta(g_1, g_3) \: - \: \theta(g_3, g_1) \, 
\right]
\, \right)
\end{eqnarray*}
Since the gauge fields $\theta(g_i,g_j) \propto dx^{11}$,
and only the path $x \rightarrow g_3 \cdot x$ winds around the 
eleven-dimensional circle, the factors corresponding to the 
paths $x \rightarrow g_1 \cdot x$ and $x \rightarrow g_2 \cdot x$
are both trivial.  Thus, the phase factor above simplifies to merely
\begin{displaymath}
\exp \left( - \int_x^{g_3 x} \left[ \theta(g_1,g_2) \: - \:
\theta(g_2,g_1) \right] \right)
\: = \: \frac{\omega^{g_2,g_1} }{ \omega^{g_1,g_2} }
\end{displaymath}
which is precisely the IIA $T^2$ twisted sector phase factor for
discrete torsion.  Thus, we have checked that our description of
the M theory dual of IIA discrete torsion is correct, by verifying
that wrapped membranes reproduce the same twisted sector phase
factor as IIA strings.

Finally, let us check that the M theory dual of our momentum shift 
degrees of freedom reproduces the IIA twisted sector phase factors,
in the same manner.  In this case, all the $\theta(g_1,g_2)$ and
$\omega^{g_1,g_2,g_3}$ are trivial; the only nontrivial factors
in~(\ref{cfieldphase}) are given by the ${\cal B}(g) = \Lambda(g) 
\wedge d x^{11}$ factors.  Thus, the phase factor~(\ref{cfieldphase})
simplifies to become merely
\begin{displaymath}
 \exp \, \left(
\, \int_1 {\cal B}(g_1) \: + \:
\int_2 {\cal B}(g_2) \: + \: \int_3 {\cal B}(g_3) \, \right)
\end{displaymath}
We wrap the membrane on the eleven-dimensional circle as before,
so $g_3 = 1$ and the path $x \rightarrow g_3 \cdot x$ wraps once
around the eleven-dimensional circle.  Looking at figure~\ref{t3phase},
we see that side~3 does not wrap around the eleven-dimensional circle
at all, so there is no contribution to the phase factor from ${\cal B}(g_3)$.
Side~1 wraps once around the eleven-dimensional circle, and also
around the path $x \rightarrow g_2 \cdot x$, hence 
\begin{displaymath}
\int_1 {\cal B}(g_1) \: = \: \int_x^{g_2 x} \Lambda(g_1)
\end{displaymath}
Similarly, since side~2 wraps once around the eleven-dimensional circle
and also around the path $x \rightarrow g_1 \cdot x$, we have
\begin{displaymath}
\int_2 {\cal B}(g_2) \: = \: - \: \int_x^{g_1 x} \Lambda(g_2)
\end{displaymath}
(where the relative sign arises from relative orientations).
Thus, the total phase factor~(\ref{cfieldphase}) finally reduces,
for this wrapped membrane, to
\begin{displaymath}
\exp\left( \int_x^{g_2 x} \Lambda(g_1) \: - \:
\int_x^{g_1 x} \Lambda(g_2) \right)
\end{displaymath}
Thus, as advertised, we have again recovered the IIA twisted sector
phase factor (this time for shift orbifolds) from the $C$ field phase factor on a wrapped membrane.

Thus, both IIA discrete torsion as well as the momentum shift degrees of freedom
discussed in this paper are encoded in M theory as some obscure
orbifold group actions on the $C$ field, orbifold group actions closely
analogous to the shift orbifolds we have discussed in this 
paper for the $B$ field.

\section{Conclusions} \label{concl}

In this paper we have done the following two things:
\begin{itemize}
\item We have clarified the physical meaning of
non-discrete-torsion orbifold group actions on $B$ fields,
by showing those additional orbifold group actions correspond to 
momentum-shift phases \cite{dgh,katrin}.
The answer was implicit in our previous work \cite{medt}; we have
hopefully clarified the matter here.
Also, in principle our methods allow us to make sense of momentum-shift
phases for orbifolds of non-simply-connected spaces other than tori,
though we have not pursued that possibility.
\item We have explicitly described the M-theory dual of IIA discrete torsion,
which is slightly obscure as M theory has no $B$ field.
\end{itemize}

\section{Acknowledgements}

We would like to thank J.~Evslin, S.~Katz, and 
especially K.~Wendland for useful conversations.

\end{document}